\documentclass[conference]{IEEEtran}
\IEEEoverridecommandlockouts
\usepackage{cite}
\usepackage{amsmath,amssymb,amsfonts}
\usepackage{algorithmic}
\usepackage{graphicx}
\usepackage{textcomp}
\usepackage{xcolor}

\usepackage{balance}

\usepackage[english]{babel}
\usepackage{times}
\usepackage{xspace}
\usepackage{ifthen}
\usepackage{amsfonts}
\usepackage{amssymb}
\usepackage{graphicx}
\usepackage{rotating}
\usepackage{xfrac}
\usepackage{array}
\usepackage{color}
\usepackage{xfrac}
\usepackage{fancyvrb}
\usepackage{multirow}

\usepackage{booktabs}
\usepackage{listings,algorithm,algorithmic}

\definecolor{verylightgray}{gray}{0.98}
\usepackage{listings}
\usepackage{courier}
\usepackage[framemethod=TikZ]{mdframed}

\usepackage{booktabs}
\usepackage{amsmath}
\usepackage{algorithm}

\usepackage{caption}
\usepackage{subcaption}

\usepackage{listings}
\usepackage{color}

\definecolor{dkgreen}{rgb}{0,0.6,0}
\definecolor{gray}{rgb}{0.5,0.5,0.5}
\definecolor{mauve}{rgb}{0.58,0,0.82}

\graphicspath{{figures/}}
\def\BibTeX{{\rm B\kern-.05em{\sc i\kern-.025em b}\kern-.08em
    T\kern-.1667em\lower.7ex\hbox{E}\kern-.125emX}}

\newboolean{showcomments}
\setboolean{showcomments}{false}
\ifthenelse{\boolean{showcomments}}{%
   \newcommand{\bnote}[2]{
    \fbox{\bfseries\sffamily \color{blue}{#1}}
    {\sffamily$\blacktriangleright$\color{red}{#2}$\blacktriangleleft$}}
}{%
   \newcommand{\bnote}[2]{}
}

\newcommand{\eg}{e.g.,\xspace}
\newcommand{\etal}{\emph{et al.}\xspace}

\newcommand{\secref}[1]{Section~\ref{#1}\xspace}

\newcommand{\figref}[1]{Figure~\ref{#1}\xspace}

\newcommand{\tabref}[1]{Table~\ref{#1}\xspace}

\newcounter{rq}
\newcounter{q}
\newenvironment{Marking}{
	\vspace{+0.1cm}
	\begin{mdframed}[style=MyWhiteBlackFrame]
}
	{\end{mdframed}}

\newcounter{f}
\newcounter{deflist}
\newcounter{mydef}
\newcommand{\myparagraph}[1]{\noindent\textbf{#1}}

\newcommand{\GROUM}{\emph{groum}}
\newcommand{\GROUMiner}{\emph{GrouMiner}}

\newtheorem{Definition}{Definition}
\newcommand{\tool}{\textsc{APImmune}\xspace}
\newcommand{\code}[1]{{\footnotesize\texttt{#1}}}
\newcommand{\Iterator}{\code{Iterator}\xspace}
\newcommand{\crypto}{\code{crypto}\xspace}
\newcommand{\http}{\code{http}\xspace}

\begin{document}
\title{API Misuse Detection \\An Immune System inspired Approach}

\author{\IEEEauthorblockN{Maxime Gallais-Jimenez}
\IEEEauthorblockA{\textit{Universit\'e de Montr\'eal,} \\
Montr\'eal, Canada \\
maxime.gallais-jimenez@umontreal.ca}
\and
\IEEEauthorblockN{Hoan A. Nguyen}
\IEEEauthorblockA{\textit{Iowa State University,} \\
Iowa, USA \\
hoan@iastate.edu}
\and
\IEEEauthorblockN{Mohamed Aymen Saied}
\IEEEauthorblockA{\textit{Laval University} \\
Québec , Canada \\
mohamed-aymen.saied@ift.ulaval.ca}
\and
\and
\and
\IEEEauthorblockN{Tien N. Nguyen}
\IEEEauthorblockA{\textit{University of Texas at Dallas} \\
Texas, USA \\
tien.n.nguyen@utdallas.edu}
\and
\and
\and
\and

\IEEEauthorblockN{Houari Sahraoui}
\IEEEauthorblockA{\textit{Universit\'e de Montr\'eal,} \\
Montr\'eal, Canada \\
sahraouh@iro.umontreal.ca}

}

\maketitle

\begin{abstract}
APIs are essential ingredients for developing complex software systems. However, they are difficult to learn and to use. Thus, developers may misuse them, which results in various types of issues. In this paper, we explore the use of a bio-inspired approach (artificial immune system) to detect API misuses in client code. We built \tool{}, a novel API misuse detector. We collect normal usages of a given APIs from the set of client programs using the APIs, especially after some API usages were fixed in those programs. The normal API usages are considered as normal body cells. We transform them into normal-usage signatures. Then, artificial detectors are randomly generated by generating artificial deviations from these usages with the objective of being different from the normal usage signatures.  The generated detectors have the ability to detect risky uses of APIs exactly as the immune system detects foreign cells of the organism. Moreover, for the detection purpose, only the artificial detectors are necessary, without the need to disclose the code used to generate them. Our approach was evaluated on the misuses dataset of three APIs as well as on known misuses from a state of the art APIs misuses benchmarking dataset.  \tool{} was also compared to four state-of-the-art API misuse detection tools. The results show that \tool{} has good detection accuracy and performance, and it can complement pattern-based tools for uncommon misuses detection.
\end{abstract}


\section{Introduction}

Nowadays, software libraries and their Application Programming
Interfaces (APIs) are essential ingredients for the development of
complex software systems \cite{vayghan2018deploying, almarimi2019web,  vayghan2019kubernetes, vayghan2019microservice, saidani2019towards}. They are supposed to provide tested and
proven reusable functionality at a low cost \cite{shatnawi2018identifying}. Yet, using APIs is not
always easy due to their complexity and often incomplete
documentation \cite{saied2015observational,benomar2015detection,shatnawi2018identifying2}. Developers may misuse them, which results in
faults difficult to debug. Preventing API misuses is not always
possible due to their complexity and the many different ways they can
be used. An alternative approach is to detect API misuses. The
existing detection approaches for API misuses can be broadly
classified into two categories: explicit and implicit detection.

Several approaches fall into the explicit detection category in
which the specifications of correct API usages are determined either
by the API designers or developers. The specifications are then used
in the detection process where a detector assesses if a given usage is
valid with respect to API specifications. However, not all libraries
are well-documented for correct usages because the number of correct
usages can be large. 

Manually writing and
maintaining specifications over time is then
challenging~\cite{dagenais-icse12,dagenais-fse10}.

To avoid the need of writing specifications or explicitly
enumerating all the potential usages, several approaches follow
implicit detection in which the correct API usages are not specified
before the detection. These approaches rely on the {\em consensus
  principle} in which an instance of an API usage is considered as a
misuse if it deviates far from the frequently used ones (often called
API usage {\em patterns}) \cite{uddin2012temporal,saied2015mining, saied2015could,zhong2009mapo, saied2015visualization, saied2016cooperative, saied2015could2, saied2016automated}.
%Tien
While several researchers have been proposing a large number of mining
techniques on API misuse detection, API misuses remain a problem in
practice~\cite{LHXRM16,ABFKMS16}.show that existing API misuse detectors have suffered high false positives due to an important issue with the use of thresholds on frequent usages (patterns) and on the deviations from those patterns. Specifically, the detectors often failed to detect misuses when they cannot relate misuses to the respective patterns because the differences between them exceed pre-defined thresholds. 

To address the challenges in establishing the distinctions between
correct usages and misuses or in spending efforts in writing
specifications, we explore an idea from biology.
A similar detection phenomenon is one from the biological immune
system (BIS). The goal of this detection is to decide if an element is
a normal body cell (\emph{self}) or a threat that can be an antigen, e.g.,
bacteria, viruses, and parasites, or a malfunctioning cell, e.g.,
cancerous cells (\emph{non-self}). The detection is done as the BIS produces
among others T-cells that are created randomly and kept only if they
do no match normal body cells. The number of self-elements (body
cells) is very high, and the number of non-self elements is
potentially infinite. Still, except for very few cases, 
the detection with the T-cells works very well. A key
benefit of this detection mechanism in BIS is to minimize the
false-positive rates, i.e., minimize the mis-identification of normal
cells as non-selfs. 

In this work, we explore that BIS-inspired idea to build {\tool}, a
novel API misuse detector. The normal API usages are considered as the
normal body cells (self). We will collect normal usages of the given
APIs from the set of client programs correctly using the APIs.
For those API usages, {\tool} extracts features to be used as usage
signatures. Like for the BIS, artificial detectors (equivalent of
T-cells) are randomly generated with the objective of being different
from the normal usage signatures. The detection is performed when
{\tool} is used to detect API misuses in a given client program using
the APIs. The API usages in the client program are extracted and
compared against the detectors. Matches represent misuse
risks. With this BIS-inspired mechanism, {\tool} avoids the
establishing of the thresholds and frequencies as in the
consensus-based approaches, and avoids manual writing of API
specifications as in the explicit approaches.  Importantly, as a
consequence, as {\tool} creates the detectors by mutating the normal
usages, we expect it to improve false positive rates over the existing
approaches. Moreover, when the detectors are generated, they can be used and 
enhanced without the need of disclosing the clients' code that served for the generation.

To evaluate the
viability of our approach, we performed a preliminary study with three APIs. 
Our results show that {\tool} has good detection accuracy and performance, 
and it can complement pattern-based tools for uncommon misuses. 

The rest of the paper is organized as follows. In sections \ref{background} and  \ref{section:formulation}  we introduce the principles of the artificial immune system algorithm (AIS) and discuss the parallel between the immune system and the detection of API misuse. The different steps of our approach are described in Section~\ref{sec:approach}. Section~\ref{sec:experiments} presents the results of the preliminary evaluation while providing discussions in Section~\ref{sec:validity}, 
Section~\ref{sec:related} presents the closest related work and the novelty of our approach with respect to it. Finally,we conclude and suggest future work in Section~\ref{sec:conclusion}.

\section{Background}
\label{background}

In this section, we present the background on API usages and misuses
as well as on the principles of the artificial immune system (AIS), a
simplified abstraction of the biological immune system (BIS).

\subsection{API Misuses}

Developers use the functionality of libraries~via Application
Programming Interfaces~(API) to access the classes, methods, and
fields that make up the APIs. Software libraries can be used in
different ways. API specifications are the conditions on the usages of
those API elements that a program needs to follow for the libraries to
work properly \cite{saied2018improving,huppe2017mining}. For example, in Java Development Kit (JDK), one could
instantiate a \code{BufferedReader} object for reading the data from a
buffer, and then close the resource to guarantee data integrity.

However, not all of the usages are well documented in the official
documentation and programming
guides~\cite{DBLP:conf/icse/Duala-EkokoR12}. That leads to misunderstanding
and, thus, incorrect usages of the APIs that violate their specifications.
Those violations are called {\em API misuses}.
For example, in Listing~\ref{example}, the resource
\textit{BufferedReader} is instantiated (line 2), used to read each
line (line 4) but if not closed at the end, it will be a classic example of
\textit{BufferedReader} misuse. This misuse is an instance of one of
the 13 types of common misuses, identified by the authors of the
benchmark \emph{MuBench}~\cite{MuBench}, i.e., missing call.

\begin{lstlisting}[
  %aboveskip=10mm,
  basicstyle=\rm\footnotesize, 
  %belowskip=10mm,
  breaklines=true, 
  caption={\textit{BufferedReader usage example}}, 
  captionpos=b, 
  %emph={hasNext, next, remove, Iterator, iterator},
  emphstyle={\color{red}},
  float=*h, 
  frame=lines, 
  framerule=2pt, 
  framesep=5mm, 
  keywordstyle=\color{blue}, 
  label={example}, 
  language=java, 
  linewidth=0.9\linewidth, 
  numbers=left, 
	numbersep=5pt,
  numberstyle=\tiny, 
  rulecolor={\color[gray]{0.6}}, 
  rulesep=1mm, 
  rulesepcolor={\color[gray]{0.9}}, 
  stringstyle=\color{teal}, 
  xleftmargin=0mm, 
  xrightmargin=5mm, 
	tabsize=2,
]
StringBuffer strbuf = new StringBuffer();
BufferedReader in = new BufferedReader(new FileReader(file));
String str;
while((str = in.readLine()) != null) {
	strbuf.append(str);
}
if(strbuf.length() > 0) {
	outMessage(strbuf.toString());
}
in.close();
\end{lstlisting}

\subsection{Artificial Immune System Detection}
A detailed presentation of the biological immune system and Artificial Immune Systems  can be found in \cite{immunology2002richard,castro2002artificial}. 
Let us summarize the principles of the artificial immune system algorithm (AIS) that interests us for our work.

To protect the organism from potential pathogens, the immune system
follows a 3-step cycle: (1) discovery, (2) identification and (3)
elimination. The discovery step detects potential pathogenes, such as
viruses and bacteria. When such an element is detected, the
identification step is responsible for checking if the identified
element is known (immune memory). Finally, in the elimination step,
the adequate response is selected depending on the identification
step.  Discovery is the phase that interests us in particular as we are concerned with the detection of API misuses. Therefore, we explain its principle in the following paragraphs.

There is no central organ that fully controls the immune system. In-stead, detectors wander in the body searching for harmful elements. Any element that can be recognised by the immune system is called an antigen. The cells that originally belong to our body and are harmless to its functioning are termed self (for self antigens) while the disease-causing elements are named non-self (for non-self antigens). The immune system classifies cells that are present in the body as self and non-self cells.

The immune system produces a large number of randomly created detectors T-cells that can be used to detect non-self elements. The T-cells are created randomly and exposed to normal cells. If a T-cell matches a normal cell, it is removed from the repertoire of immune cells to avoid the body attacking itself. This is called negative selection. When using the AIS metaphor, it is not possible to create a large number of T-cells. Hence, it is important to ensure a maximum coverage while keeping minimal the number of T-cells

The next important notion of an AIS is the affinity computation
between a detector and an encountered cell. The affinity is a
similarity function that assesses if the encountered cell
belongs to self or not.

\begin{figure}[ht]
\centering
\includegraphics[width=0.9\columnwidth]{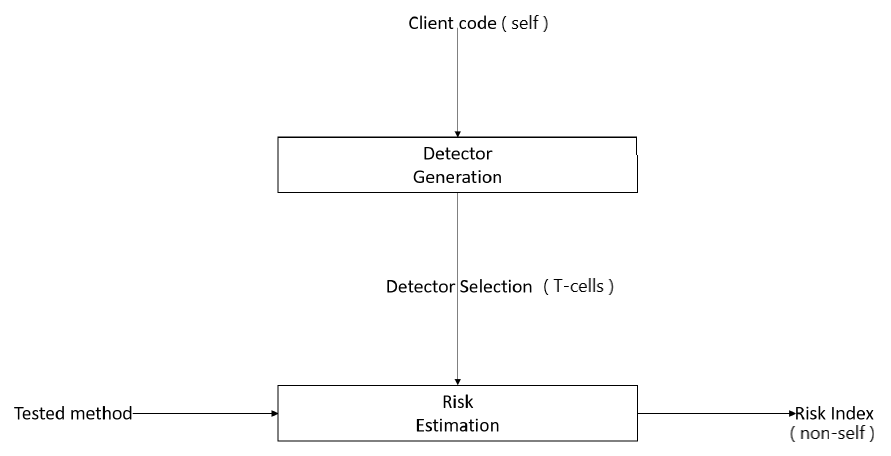}
\caption{AIS concepts applied to the API misuse detection}
\label{AIS}
\end{figure}

Figure \ref{AIS} gives a simplified overview of how the presented AIS concepts will be used in our approach. The normal API usages are considered as the normal body cells (equivalent of the self). We will collect normal usages of the given APIs from the set of client programs correctly using the APIs.  Artificial detectors (equivalent of the T-cells) will be randomly generated with the objective of being different from the normal usage of APIs. The detection is will be performed on tested methods using the APIs, that will be compared against the detectors to estimate the misuse risks. The risk index will be used to identify the  API misuse (equivalent of the non-self).

\section{BIS-inspired API Misuse Detection Formulation}
\label{section:formulation}

In this section, we present our formulation of the API misuse
detection problem by the adaptation of the BIS.

\begin{Definition}[API Usage]
{\em An API usage is a fragment of client code that involves the API classes and/or
methods for a given library.}
\end{Definition}

The code fragment in Listing~\ref{example} is an example of usage of
APIs in JDK library with classes
such as \code{StringBuffer}, \code{BufferredReader}, and \code{FileReader},
and method calls such as \code{BufferedReader.readLine()},
\code{StringBuffer.append(...)}, and \code{StringBuffer.length()}.

\begin{figure*}[h]
\includegraphics[height=4.75in, width=6.25in]{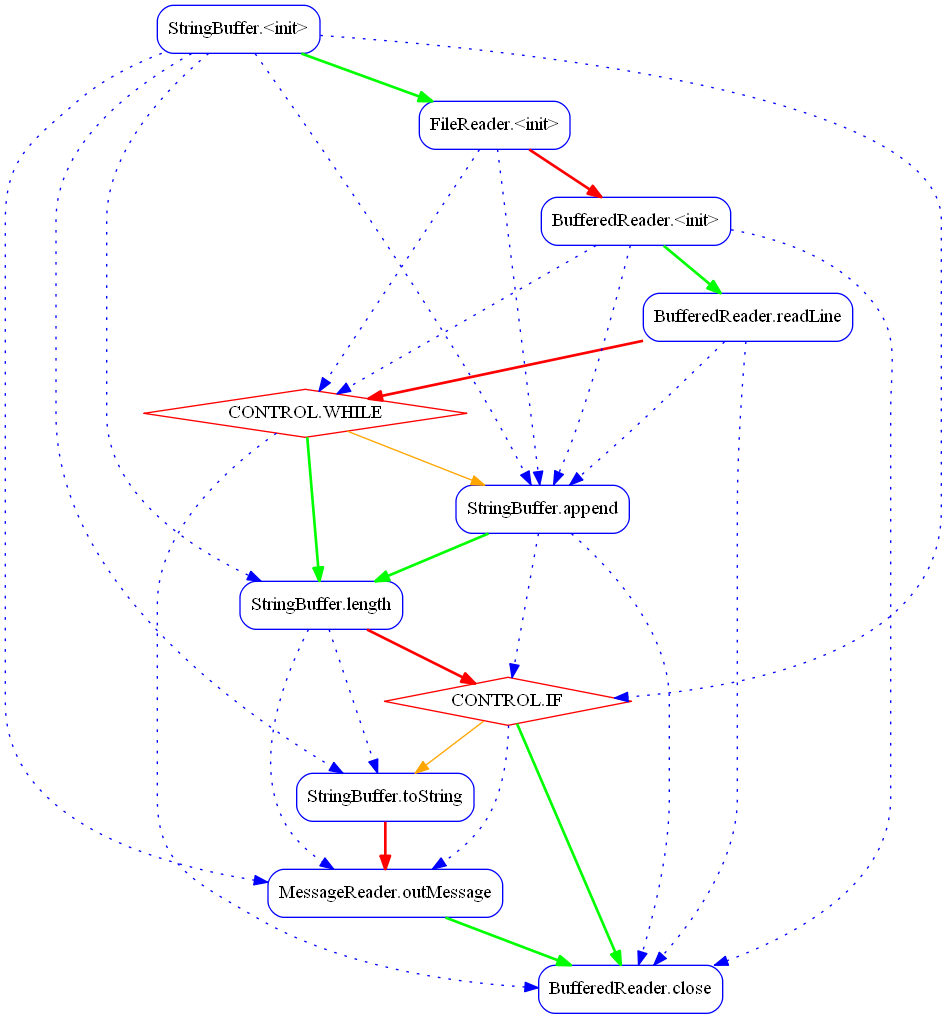}
\caption{Graph-based API Usage Representation for Listing~\ref{example}}
\label{Groum1}
\end{figure*}

When a client method is considered as an API usage, not all its statements
are relevant to this usage. To capture the specific usage, we extract a \GROUM{},
a graph-based representation for the API usage~\cite{Grouminer}.

\begin{Definition}[API Usage Graph]
{\em An API Usage Graph (GROUM) is a directed graph representing an API
usage in which nodes represent API objects constructor calls, method
calls, field accesses, and branching points of the control structures,
and edges represent temporal usage orders and data dependencies among
them.}
\end{Definition}

For example, the API usage of Listing~\ref{example} will lead to
the \GROUM{} shown in Figure~\ref{Groum1}.

As mentioned in Section~\ref{background}, in a BIS, the notion of Self
refers to what is normal by contrast to what is risky. In the API misuse
detection problem, the Self is the correct use of the API.

\begin{Definition}[API Usage as Self]
{\em The self for an API is a set of API usage methods that are known to
be correct.}
\end{Definition}

The Self for an API can be extracted from the set
of client programs that correctly use the APIs. Identifying
correct usages can be done manually based on history data
(\eg issue tracking systems).
In our experiments (see Section~\ref{sec:experiments}), we used an
automated approach to extract the Self. We included in this set
versions of methods using the API that were fixed after a bug was declared
in them. We retain a fixed version if its \GROUM{} differs from one of the
corresponding buggy version.

Another important concept in BIS is the mutations of T-cells to
detect the anomaly cells. Roughly speaking, the immune system generates
T-cells from the stem cells and keep only those that do not match body
cells (negative selection). When transposing the principle of the
negative selection to the detection of API misuses, it is necessary
to generate detectors (equivalent to T-cells) by randomly mutate
the Self API usages. As the goal is to generate a limited number
of detectors with a maximum coverage of the deviations from the Self,
the generation process can be viewed as a {\em multi-objective
optimization} problem. If we fix the number of detectors for performance
consideration, the goal is, therefore, to find the set of detectors that
deviate from the normal use situations in the Self,
but also are as much as possible different from one another to avoid duplications.

\begin{Definition}[Detector Generation]
{\em The generation process aims to produce a fixed number of detectors that
deviate from the normal API usages in correct client programs, and that are
different as much as possible from one another.}
\end{Definition}

\begin{Definition}[A Mutation]
{\em A mutation of an API usage $u$ is the \GROUM{}
of $u$ after the application of a mutation operation.}
\end{Definition}

\begin{Definition}[Mutation Operators]
{\em Mutation operators are respectively: adding an edge to or removing
an edge from the \GROUM{}, adding a
node (a random API method call), removing a node, replacing a node
(changing a method call by another), moving (changing the
position of) a node in the \GROUM{}, and adding, removing and moving
an exception.}
\end{Definition}

%Tien
For example, a mutation of the correct usage from Figure~\ref{Groum1}
is the graph without the node \code{BufferedReader.close} and the
inducing edges to that node.

The next important notion to instantiate in our formulation is the
affinity computation that assesses the similarity between a T-cell
(detector) and an encountered cell (an evaluated client method) to
determine if the cell is part of the Self or not.

\begin{Definition}[Affinity of APIs]
{\em In the API misuse detection context, the similarity between a
  detector graph an  API usage in a client program is measured by
  the graph editing distance between the respective groums.}
\end{Definition}

The basic mapping between a BIS and our detection problem is not enough
to fully tackle the complexity of the API misuse problem.
In this section, we present the additional features that complement
the BIS-inspired detection.

The first feature is the clustering of the API usages before the
generation of the detectors. Many APIs can be used in different ways by
different clients. To ensure addressing all these different usage variations,
we apply a clustering process to the API usages in the Self to
group similar API usages in clusters. Each cluster corresponds to a usage
scenario. Then the generation process will take into account the
representativeness of the generated detectors with respect to the obtained
clusters.

The second feature is to replace the boolean detection results (Self/non-Self)
by a risk
score that allows to rank the evaluated methods according to the estimated
risk. This helps client-program developers dedicating their available effort
to the methods with the higher risk  scores.

\section{{\tool}: BIS-inspired API Misuse Detection}
\label{sec:approach}

This section presents our algorithms to realize {\tool}, an
BIS-inspired API misuse detection tool.

\begin{figure*}[h]
\centering
\includegraphics[width=5.25in]{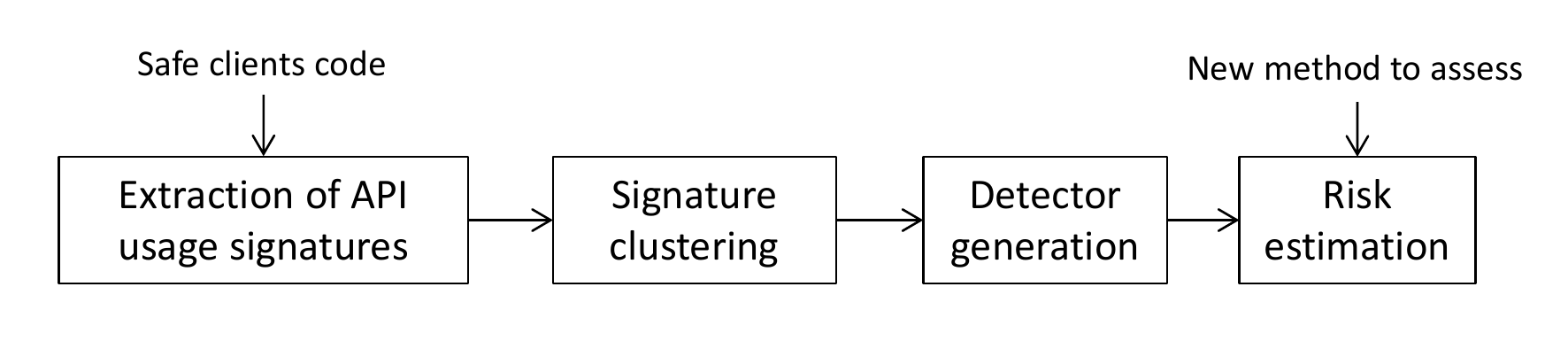}
\caption{{\tool} Architecture: BIS-inspired API Misuse Detection}
\label{fig:Approach}
\end{figure*}

Our approach to detect API misuses is depicted in
Figure~\ref{fig:Approach}. We begin with the extraction of usage
signatures (groums) that represent the usage scenarios from the methods of a
given client code corpus (Self). As the API can be used in different ways,
the next step is to cluster the signatures depending on which API
methods are involved. Then, starting from the clusters, a
set of detectors is generated. The final step is the actual detection, in
which all the generated detectors are used to assess the misuse risk
of each new client method. Note that the obtained detectors will have
their own independent life cycles. They can be reused/shared, enhanced with
new detectors when new safe clients are considered and destroyed if
they detect false positive(s).

\subsection{Usage Signature Extraction}
\begin{figure}[h]
\centering
\includegraphics[width=0.6\columnwidth]{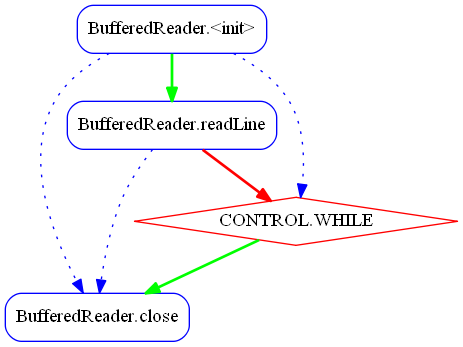}
\caption{Usage Signature Example}
\label{Groum2}
\end{figure}

The goal of this step is to produce, for each method in the safe clients' code, a signature that captures the API usage independently from the client behavior. To this end, we use the tool \GROUMiner~\cite{Grouminer} to extract a \GROUM{} as defined in Section~\ref{section:formulation}. The initial \GROUM{} contains all the elements in the considered-method body. Then, this \GROUM{} is pruned by removing all the nodes and edges that are not concerned with the API calls. Let us consider again the code fragment of Listing~\ref{example}, the initial \GROUM{} is the one of Figure~\ref{Groum1}. Now if only the BufferedReader class is considered as part of the API, the pruning process will produce the \GROUM{} depicted in Figure~\ref{Groum2}.

\subsection{Signature Clustering}
\begin{figure*}[h]
\centering
\includegraphics[height=2.75in, width=6in]{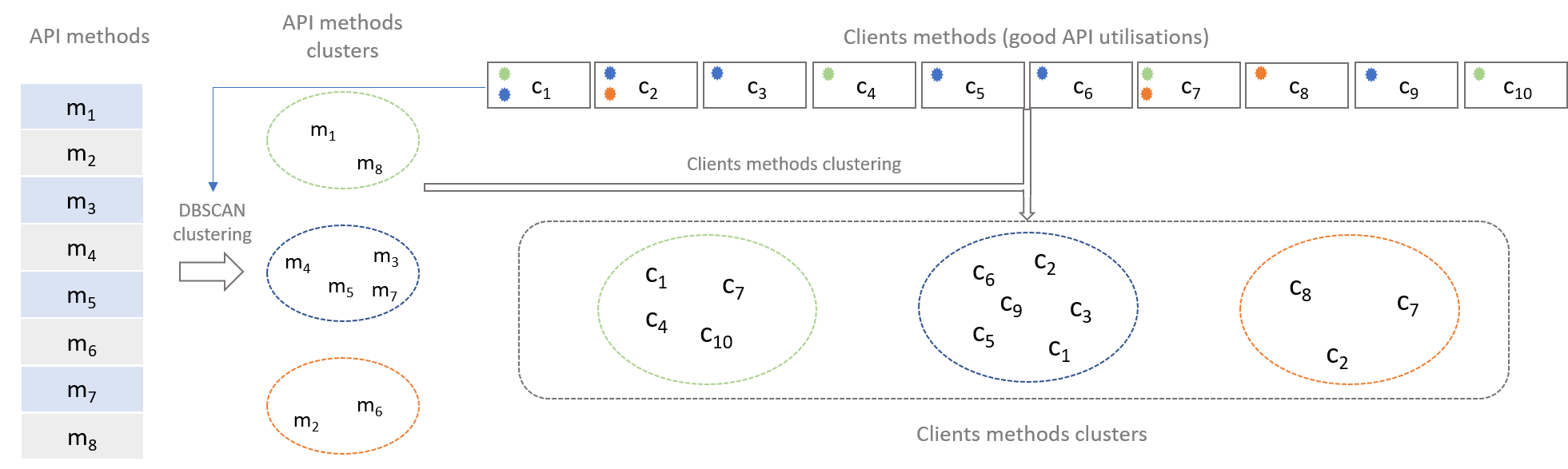}
\caption{Signature clustering}
\label{fig:clustering}
\end{figure*}

An API can offer different functionalities and then expose different sets of methods to use them. It is, then, important to target different families of misuses without enumerating them explicitly. A good way to handle this variety of API usages is to identify similar usage scenarios. In this context, the second step of our approach is to derive clusters of usage-signatures, which define families of API usage scenarios. The clustering allows targeting different usage scenarios during the detector generation and taking into account those that are not very common. Moreover, if the detectors are generated without clustering, redundant detectors will be derived for similar usage scenarios that were not clustered.
Figure~\ref{fig:clustering} shows the clustering process.

In a first step, We cluster API methods that are co-used together by the client methods. To this end, we use DBSCAN, a density-based clustering algorithm  ~\cite{Ester96adensity-based}. DBSCAN constructs clusters of API methods by grouping those that are close to each other  (i.e., similar methods) form a dense region. Two API methods are close to each other (short distance) if they have a high co-occurrence frequency, thus, they will share a large set of common usage.  Moreover, with DBSCAN we don't need to specify the number of clusters, and DBSCAN is also very robust to outliers which in our case will occur for utility methods that are frequently co-used with domain-specific methods.
The algorithm has two parameters: The first parameter is the minimum number of methods in a cluster. We set it at two so that a cluster includes at least two methods of the  API. The second parameter is epsilon, the maximum distance within which two points can be considered as neighbor, each to other. In other words, epsilon value controls the minimal density that a clustered region can have. The shorter is the distance between methods within a cluster the denser is the cluster. We set it at 0.8 to minimize the noisy points, i.e., two methods are clustered together if they share at least 20\% of their client calling methods.

In a second step, we derive the families of API usage scenarios. For each API-method cluster inferred in the first step, we identify its corresponding client methods, i.e., the client methods using the API methods in the cluster.

\begin{figure*}[p]
\centering
\includegraphics[height=3.5in, width=6in]{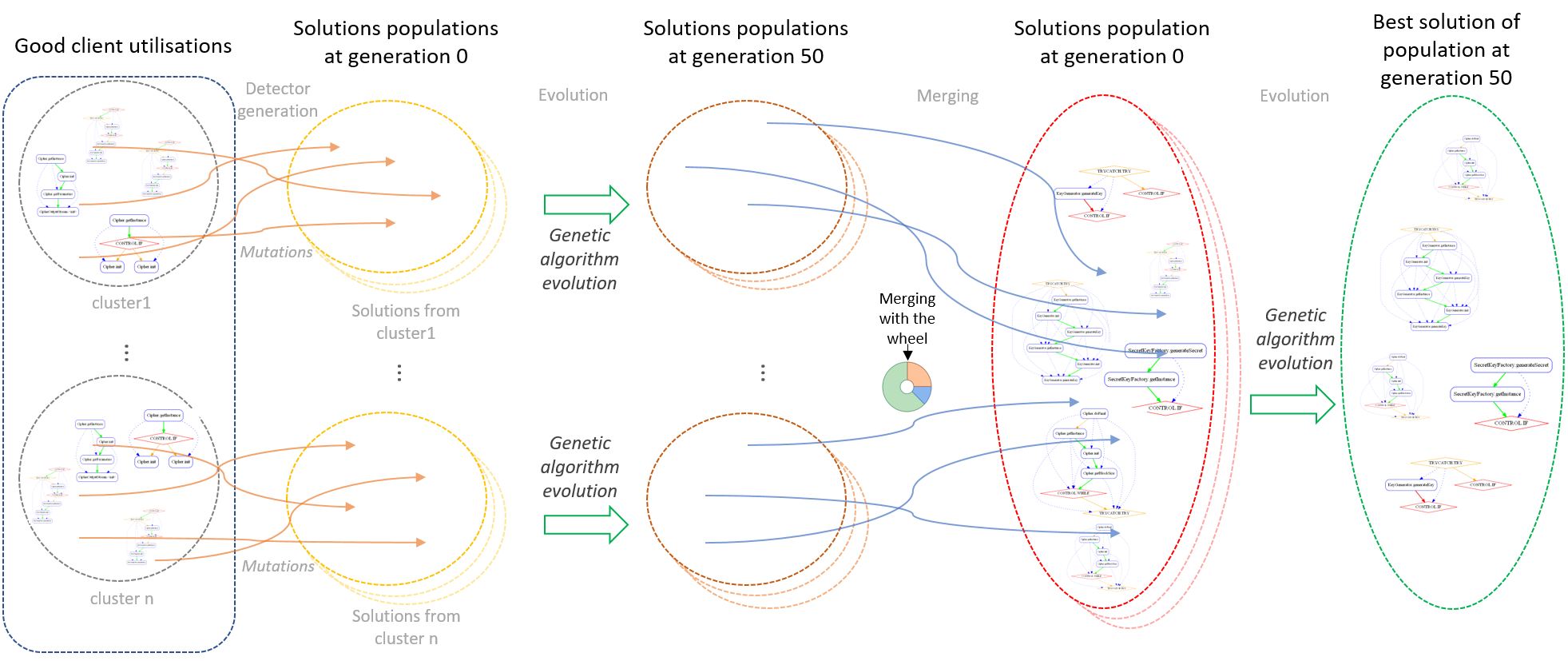}
\caption{Detectors generation process with clustering. Process with merging after evolution}
\label{fig:evolutionTop}
\end{figure*}

\begin{figure*}[p]
\centering
\includegraphics[height=3.5in, width=6in]{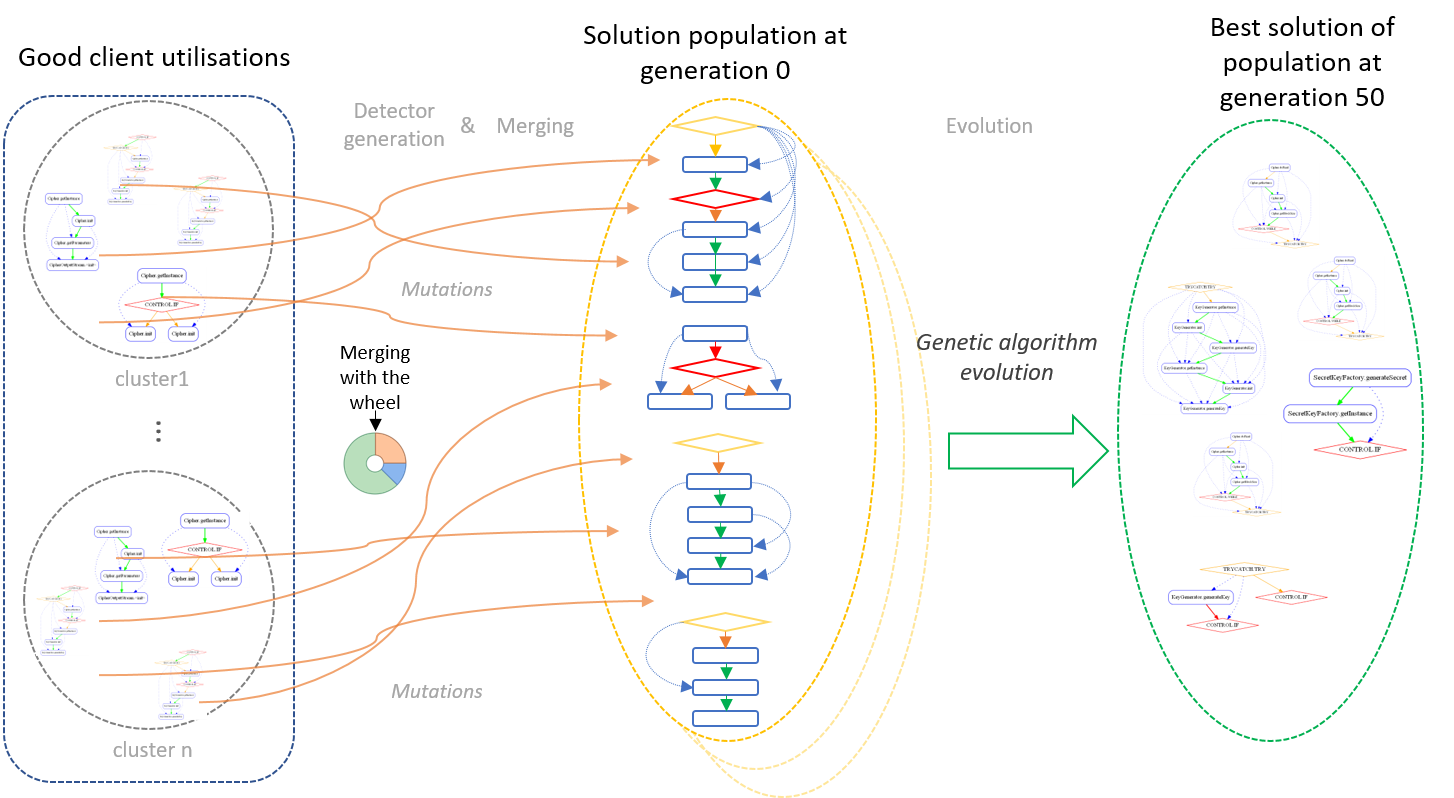}
\caption{Detectors generation process with clustering. Process with merging before evolution on the bottom}
\label{fig:evolutionbottom}
\end{figure*}

\subsection{Detector Generation}

To allow the detection of API misuses, we use a genetic algorithm to generate
detectors mimicking the T-cells. The objective is to generate a fixed number
of detectors that represent artificial signatures that are different from those of the safe code
(random alterations of the good-usage signatures).
The genetic-based generation algorithms start from a population of randomly
generated detector sets, each having a fixed size. Each set is a candidate
solution. Then the algorithm evolves these sets through a given number of generation.
In each generation, the algorithm create a new population of candidate detector sets
by modifying the detectors in the sets (production of new genetic material by the mutation operator)
and/or by combining detectors coming from different sets (recombination of
the existing genetic material by the crossover operator).

After the clustering step,  we experimented with two alternative generation process.
The first, call it \textit{parallel evolution}, consists in having a separate
detector generation process (Figure~\ref{fig:evolutionTop}). The second, call
\textit{global evolution}, uses the clusters to seed the generation process by
producing an initial set of detector that are globally refined later on
(Figure~\ref{fig:evolutionbottom}).

 \textit{Parallel evolution:} The detector generation is performed specifically
 for each cluster in a parallel mode. Detectors are generated by
 mutating client-method groums. We evolve detectors separately for each cluster,
 using a genetic algorithm, and we merge the best final solutions (set of detectors) at the end of the process. If the number of clusters is high, merging all the detectors may results in a large set of detectors. Thus, we use another optimization process to reduce the number of the merged detectors to a minimal set. To this end, we use a proportionate selection, also known as roulette wheel selection, to generate a population of random combinations of detectors with each combination having a fixed size. The probability of a detector to be included in any of the combinations is proportional to its fitness (see below for the fitness calculation). Then the generated population evolves through genetic recombination. Note that this last process does not generate new detectors. It only searches for the best combination of a fixed number of detectors.

\textit{Global evolution:}
In this alternative, the genetic algorithm is allied only once. We seed the initial population of detectors candidate sets with detectors coming from each cluster using the roulette wheel selection to have a representation of individuals conforming to the cluster size. The larger client-method groums are in a cluster, the more candidate detectors are likely to be generated from this cluster. Then, the genetic-basic generation process runs on this initial population regardless of the clusters that served to generate the initial population.

For all the alternatives, the evolution is guided by the two objectives of having a set of detectors that is different from the normal signatures while being diverse.  Additionally, the evolution of solutions is performed using genetic operators, i.e., elitism, crossover, and mutation. The details of the most important elements of the algorithm are as follows.

\textbf{Elitism}: When creating the next generation of candidate detector sets, a small subset of the current-generation sets having the highest fitness values is automatically added. The elitism ensures to keep the best solutions during the evolution.

\begin{figure*}[h]
\centering
\includegraphics[width=0.8\linewidth]{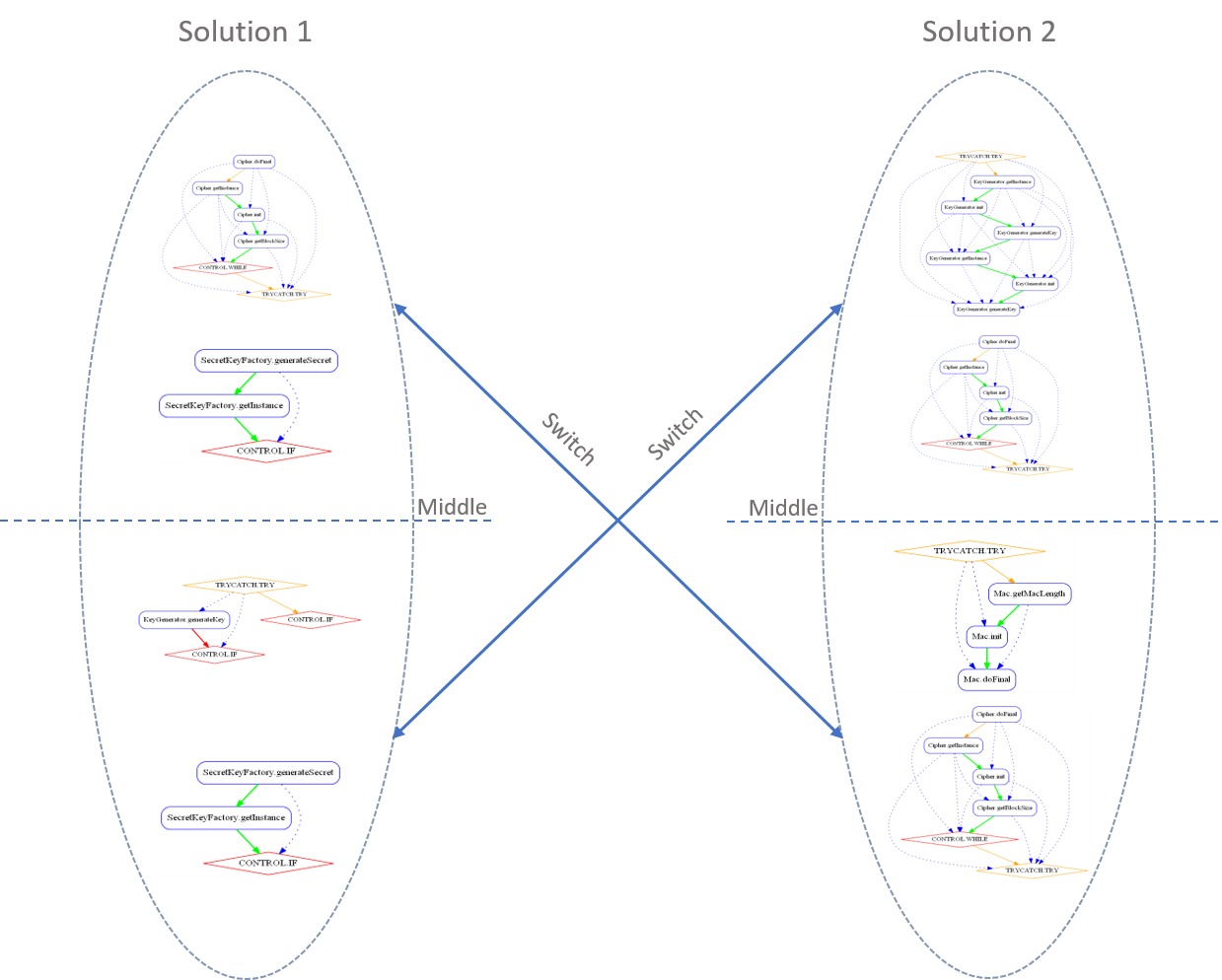}
\caption{Crossover Process}
\label{fig:crossover}
\end{figure*}

\textbf{Crossover}: After performing the elitism, the remaining slots for the
next generation are filled using the crossover between detector sets of the
current generation. The crossover consists of selecting two sets and producing two offsprings by exchanging subsets of detectors as illustrated in Figure \ref{fig:crossover}. The selection favors the fittest detector sets while keeping a certain degree of randomness. When selecting two detector sets, the crossover is performed under a probability (set to 0.9).

\begin{figure*}[h]
\centering
\includegraphics[width=0.7\linewidth]{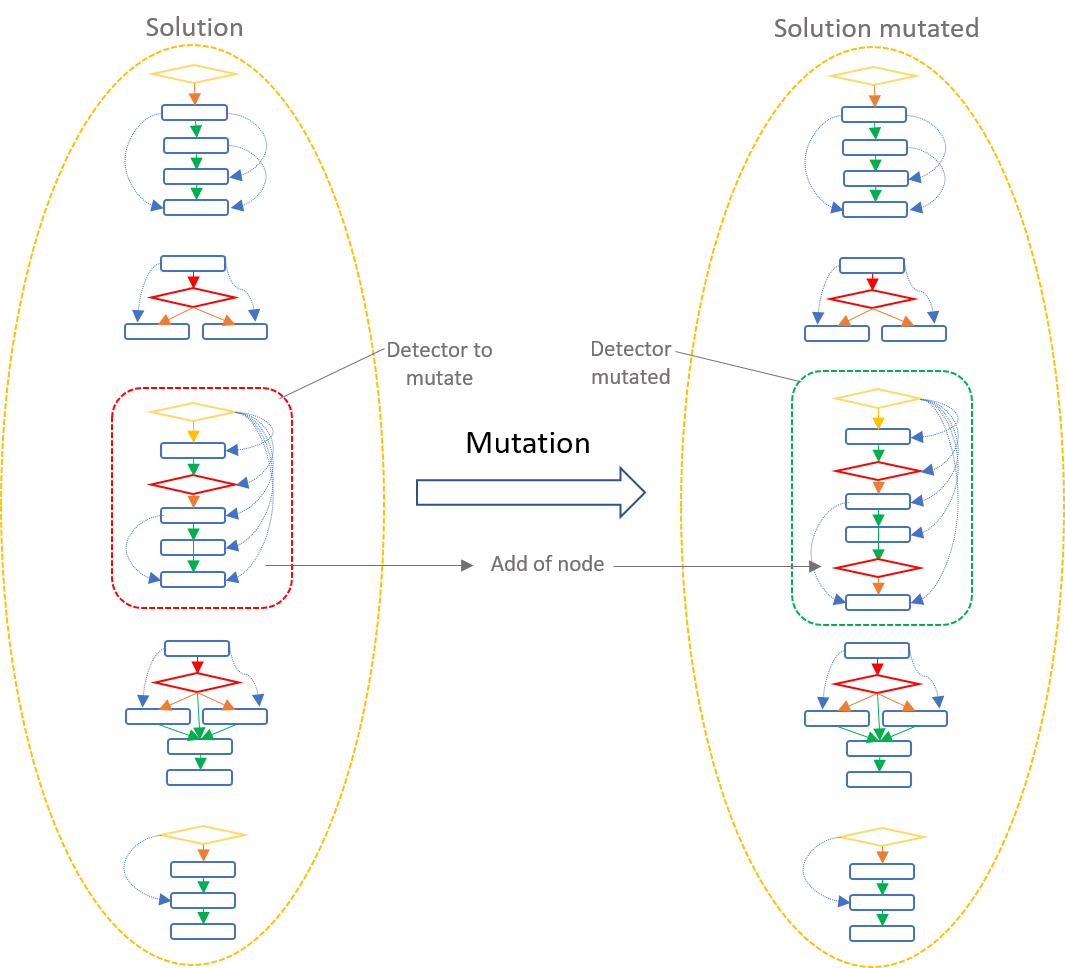}
\caption{Mutation Process}
\label{fig:mutation}
\end{figure*}

\textbf{Mutation}: After each crossover, the offsprings (or the
parents if the crossover is not performed) are candidates for mutation
with a certain probability (set to 0.2). When a decision is made to
mutate a detector set, a subset of its detectors is randomly selected
and one of the nine types of mutations is randomly selected to apply
on each of these detectors. {\tool} considers the types of mutation
operators as explained in Section~\ref{section:formulation}. Figure
\ref{fig:mutation} shows the \emph{add node} mutation.

\textbf{Fitness function}: Both the elitism and selection for the crossover use a fitness function to favor the fittest detector sets. For a detector set $S$, the fitness function is the average of the fitness scores of each detector $d_i \in S$. To ensure a maximum coverage with a limited set of detectors, the fitness function of each detector should consider two aspects: the dissimilarity with the safe usage signatures $C$ and the dissimilarity with the other detectors of $S$. Consequently, the fitness score of a detector $d_i$ is the linear combination of two dissimilarity functions:

\begin{equation}
fitness(d_i) = \alpha \cdot clientDis(d_i) + \beta \cdot detectorDis(d_i)
\end{equation}

Both client distance $clientDis(d_i)$ and detector distance $detectorDis(d_i)$ are based on the similarity function $sim(di,y)$ between two \emph{groums}: $d_i$ for the detector and $y$ for either another detector or a Self API usage. It is defined as the proportion of shared elements (nodes, edges, exceptions and control structures) between the compared \emph{groums}.

To derive $clientDis(d_i)$, we start by calculating  $minDis(d_i)$, the minimal distance between $d_i$ and any of the API usages in the Self $C$.

\begin{equation}
minDis(d_i)=1-max_{s_j \in C} (sim(d_i, s_j))
\end{equation}

To capture the fact that deviation from the normal usages are in general different but not that distant from the normal usages, we give a perfect distance score $clientDis(d_i)=1$ when the $minDis(d_i)$ is in a certain interval $[l,h]$ where $l$ is close to 0 and $h$ is a maximum tolerated deviation. We considered the interval $[0.01, 0.33]$ in our experiments. For values outside this interval, we assign to $clientDis(d_i)$ a value between $0$ and $0.75$ proportionably to how far we are from the interval. This value is 0 if $minDis(d_i)$ equals 0 or 1.

The distance with the other detectors $detectorDis(s_i)$ is calculated as follows:
\begin{equation}
detectorDis(s_i)=1-\frac{\sum_{d_k \in C,  k \neq i} sim(d_i, d_k)}{|C|-1}
\end{equation}

Regardless of detector generation alternative, the best detector set $R$ is used in the future to assess new client code.

\subsection{Misuse Detection}
\label{sec:detection}
The actual detection consists in measuring the similarity between the signature
$a_t$ of each new client method $m_t$ with each detector $d_j \in R$ using the
$sim(a_t,d_j)$ function. For a given a $m_t$, the risk score is derived by
aggregating similarities with the individual detectors.

\begin{figure*}[h]
\centering
\includegraphics[width=0.9\linewidth]{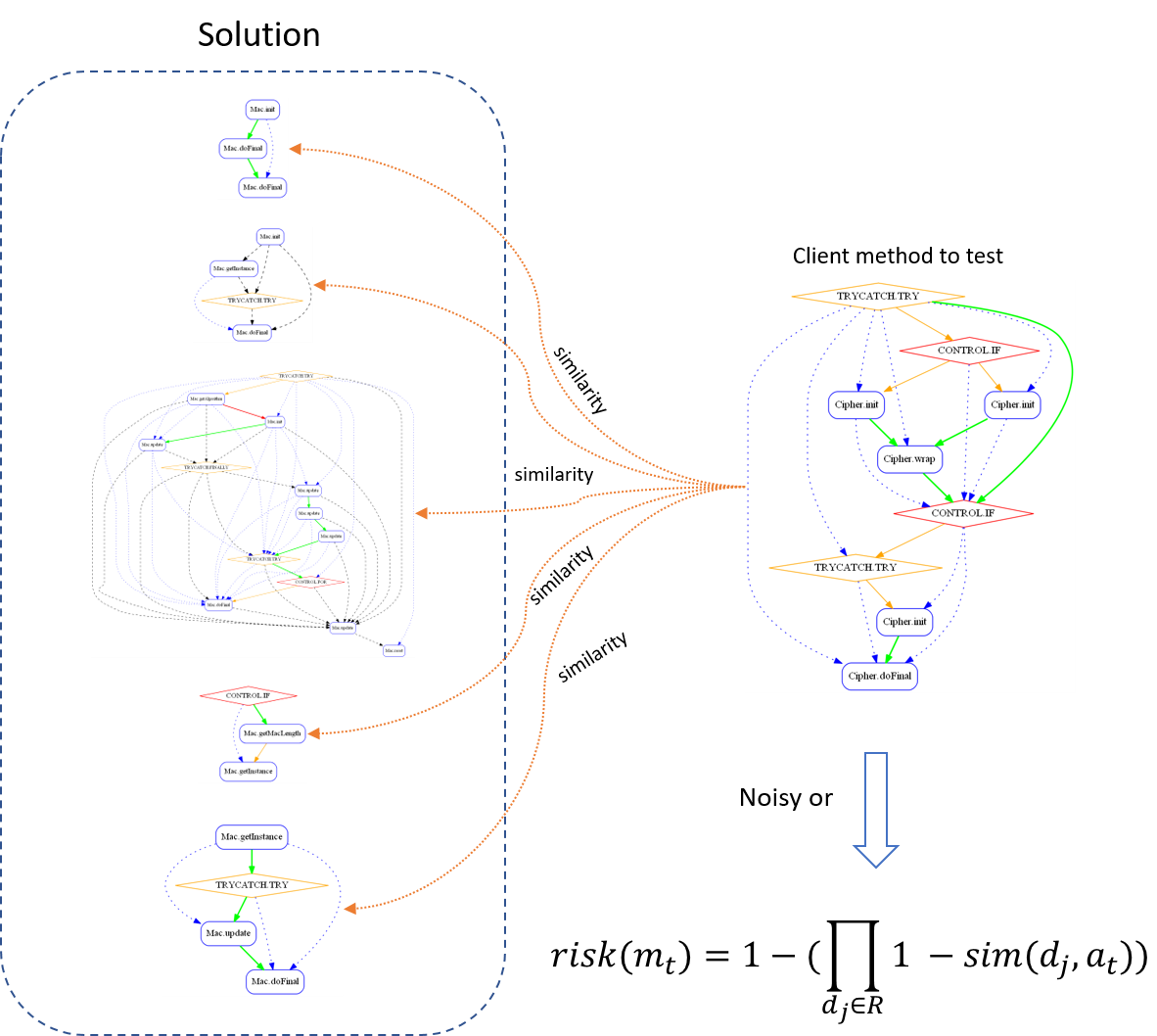}
\caption{Detectors generation process}
\label{score}
\end{figure*}

The obvious strategy is to assign to the risk
score the maximum similarity found between $m_t$ and the detectors in $R$. Alternatively,
rather than looking at the detector that best matches the
method being evaluated, we assign higher risk scores to methods that are close to multiple detectors.
The closer the method is to different detectors with high similarities, the more it will be considered at risk.
In our metaphor with the immune system, this would mean that a cell that tends to match several T-cells would
be qualified as pathogenic. To implement this idea, we use the logical function of \emph{Noisy or}.
The risque score according to the \emph{Noisy or} aggregation is calculated as follows and as illustrated in Figure~\ref{score}.

\begin{equation}
risk(m_t)=1-(\prod_{d_k \in R} 1- sim(d_j, a_t) )
\end{equation}

\section{Empirical Evaluation}
\label{sec:experiments}

\newcommand{\RQone}{\textit{What is the impact of different detector generation strategies on the misuse detection accuracy?}}
\newcommand{\RQtwo}{\textit{What is the execution cost of our approach in term of required execution time and storage?}}
\newcommand{\RQthree}{\textit{How well does our approach perform compared to existing approaches?}}

The objective of this section is to evaluate the performance of our approach
in detecting API misuses in practice and in comparison with existing techniques. We formulated the research questions of our evaluation as follows:

\begin{enumerate}[]
  \item [\textbf{RQ1.}] \RQone
  \item [\textbf{RQ2.}] \RQtwo
  \item [\textbf{RQ3.}] \RQthree
\end{enumerate}

For each experiment in this section, we present the
research question to answer, the research methodology to address it, followed by the obtained results.

\subsection{Training Data Collection}
\label{sec:data}

As explained in \secref{sec:approach}, our approach generates misuse detectors for
a specific API of interest. In our experiments, we evaluate our approach on
APIs providing different functionality from three different packages of JDK:
\code{java.util.Iterator}\footnote{https://docs.oracle.com/javase/8/docs/api/java/util/Iterator.html},
\code{javax.crypto.*}\footnote{https://docs.oracle.com/javase/8/docs/api/javax/crypto/package-summary.html} and
\code{javax.servlet.http.*}\footnote{https://docs.oracle.com/javaee/7/api/javax/servlet/http/package-summary.html}.

To train the detectors, we need to collect a large 
samples of good usages of these APIs in their client code. We did that by
looking at the changes to client methods that fixed the uses of these APIs in the histories
of open source projects hosted on GitHub. We consider the client methods before the changes (buggy ones)
as misuses and those after the changes (fixed ones) as good uses of these APIs.

In order to do that, we first collected high-quality repositories.
To eliminate toy or experimental repositories, we used GitHub APIs to search
for Java repositories that had been given at least 5 stars by individual
GitHub users. This gave us 21,745 repositories.
Then, for each repository, we identified the commits that had potentially fixed
bugs by applying string pattern matching on the contents of the commit
messages against keywords \textit{fix, bug, issue, error, problem, exception
and fail}.
For each potential fixing commit, we mapped the methods before and after the change
from the set of changed files. For each modified method, we used the abstract
syntax tree (AST) differencing algorithm in~\cite{jsync-tse11} to identify the
changed AST nodes.
If a changed method does not contain any changed AST nodes whose resolved
type is \code{java.util.Iterator} or belongs to package \code{javax.crypto}
or \code{javax.servlet.http}, we disregard it. To make sure that a changed
method actually fixes the usage of the APIs of interest, we construct and
compare the \GROUM{}s with respect to the three APIs of interest of the
methods before and after the change. If the \GROUM{}s are different, we add
the method before the fix to the set of misuses and the method after the fix
to the set of good uses of the corresponding API. We keep the misuses for the
validation experiment. The numbers of projects, fixing commits, pairs of
misuses and good uses and API \GROUM{}s generated for each API in our dataset are shown in \tabref{tab:data}.

\begin{table}
	\centering
	\caption{Statistics of the Training Dataset.}
\begin{tabular}{|l|r|r|r|r|}
\hline
API & Projects & Commits & Uses & API \GROUM{}s\\
\hline
\code{Iterator} & 832 & 1,560 & 1,833 & 3,641 \\
\code{javax.crypto} & 74 & 132 & 201 & 271 \\
\code{javax.servlet.http} & 968 & 4,672 & 6,607 & 957 \\
\hline
\end{tabular}
\label{tab:data}
\end{table}

\subsection{Accuracy Analysis}
\label{sec:eval-accuracy}

In the first experiment, we aim to answer \textbf{RQ1.} \RQone

As explained in Section \ref{sec:approach}, API misuse detectors can be generated
using different strategies. We evaluate the detection accuracy with respect to the following factors in detector generation strategies:

\begin{itemize}
  \item \textbf{Detector generation strategies:} How to generate detectors after the clustering step.
  As mentioned in Section~\ref{sec:approach}, we have two strategies to exploit the clusters
  for the detector generation: (1) \textit{parallel evolution} or (2) \textit{global evolution}.

  \item \textbf{GROUM complexity:}  \emph{groums} abstract the API usage and capture
  different aspects of the source code, which may lead to complex \emph{groums} and a
  significant overhead for detector generation and \emph{groums} comparison.
  We are interested in evaluating whether we can  achieve the same detection
  accuracy with simpler \emph{groums} in which we don't consider data dependencies.

  \item \textbf{Clustering:} One of the enhanced features introduced into our
  approach is the clustering. It helps to handle the variety of API usages, and
  it avoids the  generation of a huge yet redundant number of detectors.
  We are interested in evaluating the impact of the clustering on the detection accuracy and
  what would be the effect of omitting the clustering from our approach.

\end{itemize}

\subsubsection{Analysis method}
To answer \textbf{(RQ1)}, we performed a 10 fold cross-validation on each of the three APIs
\textit{java.util.Iterator},
\textit{javax.crypto},
and
\textit{javax.servlet.http}.
We generated misuse detectors according to three  generation configurations based on the previously mentioned variation factors: \textit{parallel vs. global
evolution}, \textit{simple vs. complex \GROUM{}s} and \textit{with vs. without clustering}. We then identified which configuration achieves the best accuracy.

For the 10 fold cross-validation, we created 10 folds that contain each 10\% of the good API usages collected (as explained in Section \ref{sec:data}). Then for each fold, we generate the detectors using the good uses from nine folds. The tenth fold, not used in the detector generation, is used in the test set. This test set is completed with the same number of bad usages, to have a balanced test set with the same number of good and bad usages.
The detectors are then applied on each good and bad use case contained in this test set, to calculate their risk score (c.f., Section~\ref{sec:detection}).
The cases are then sorted by their scores. This allows to compute the accuracy
on the top-ranked use cases since we expect the API misuse to have the highest
risk scores. We compute the accuracy as the number of true positives (misuses) over the number of considered top-k use cases. We calculate the accuracy for both top 10\% and top 30\% use cases.

\subsubsection{Results and Analysis for RQ1}
Tables \ref{tab:process5}, \ref{tab:process4}, and \ref{tab:process0} present the accuracy results for the three detector generation configurations.

\begin{table*}[htp]
	\centering
	
	\caption{Detection accuracy for \textit{Global evolution}.}
		
	\begin{tabular}{|>{\raggedright}m{2.7cm}|*{6}{m{0.5in}|}}
		\hline
		\multirow{4}{*}{}&\multicolumn{3}{c|}{Complex \emph{groums}} & \multicolumn{3}{c|}{Simple \emph{groums}} \\
		\cline{2-7}
		 & Iterator & http & crypto & Iterator & http & crypto \\
		\hline
		
		Mean accuracy 10\% data (\%)	& 52.7\%	& \textbf{70\%}	& 55\%	& 54.3\%	& \textbf{70\%}	& \textbf{60\%}	\\
		\hline
		Mean accuracy 30\% data (\%)	& 54.6\%	& 57.8\%	& 58.6\%	& 55.7\%	& 55.6\% & \textbf{65.7\%} \\
		\hline
	\end{tabular}
		
		\label{tab:process5}
\end{table*}

\begin{table*}[htp]
	\centering
	\caption{Detection accuracy for \textit{Parallel evolution}}
	\begin{tabular}{|>{\raggedright}m{2.7cm}|*{6}{m{0.5in}|}}
		\hline
		\multirow{4}{*}{}&\multicolumn{3}{c|}{Complex \emph{groums}} & \multicolumn{3}{c|}{Simple \emph{groums}} \\
		\cline{2-7}
		 & Iterator & http & crypto & Iterator & http & crypto \\
		\hline
		
		Mean accuracy 10\% data (\%)	& 52.7\%	& \textbf{60\%}	& 40\%	& 54.3\%	& 53.3\%	& 50\%	\\
		\hline
		Mean accuracy 30\% data (\%)	& 54.6\%	& 57.8\%	& 51.4\%	& 55.7\%	& 55.5\% & 50\% \\
		\hline
	\end{tabular}
		
		\label{tab:process4}
\end{table*}

\begin{table}[htp]
	\centering
	\caption{Detection accuracy for the generation without clustering and \textit{simple \emph{groums}}.}
	\begin{tabular}{|l|r|r|r|}
		\hline
		  & \multicolumn{1}{c|}{Iterator} & \multicolumn{1}{c|}{http} & \multicolumn{1}{c|}{crypto} \\
		\hline
		
		Mean accuracy 10\% data	&	54.3\%	&  50\%	& 50\%	\\
		\hline
		Mean accuracy 30\% data	&  55.7\% 	& 53.3\% & 55.7\% \\
		\hline
	\end{tabular}
	
  \label{tab:process0}
\end{table}

\textbf{\underline{Study 1.A: \textit{parallel vs. global evolution}.}}

The global evolution (Table~\ref{tab:process5}) allows achieving in general better accuracy than the parallel
evolution (Table~\ref{tab:process4}).
We conjecture that the global evolution introduces more diversity during the generation. Conversely, parallel evolution forces the individual generation of the detectors for clusters having different sizes, which in some cases limits the exploration possibilities. For example, we obtained 8 clusters of sizes 2, 2, 5, 6, 24, 44, 46, and 95 for \crypto. In the small clusters, we do not have enough examples of good usage to generate accurate detectors. In other words, if we want to generate 50 different detectors from two groums, we have to find 50 different mutations of these two groums, which is difficult when the groums are small graphs with few nodes. The same observation holds for \http, with slightly less difference between the two strategies. \http has 3 clusters of sizes 17, 51 and 61 and then has less small clusters than \crypto. Note that the results for \Iterator are exactly the same for the two strategies for the simple reason that a unique cluster was obtained (the API has only 4 methods), and then both strategies behave the same.

\textbf{\underline{Study 1.B: \textit{simple vs. complex \emph{groums}}.}}

Using simple \emph{groums} is slightly better than using complex \emph{groums}. For \Iterator, the accuracy increases by 2\% when we use simple \emph{groums}. For \crypto, the use of simple \emph{groums} increases the accuracy for both parallel and global process. In particular, for parallel process (\tabref{tab:process4}), the accuracy increases from 40\% to 50\% on the top 10\% of ranked methods. The exception is for \http where the accuracy slightly decreases for the parallel evolution but stays high (70\%) with the global evolution.

These results could be explained by the fact that the data dependency edges in the \emph{groums} are not obvious for the API usage comprehension and take an important weight during the similarity computation between a method and a detector. If we look at the \figref{Groum2} and we remove the data dependency edges in blue, we have a \emph{groums} with 7 elements instead of 10.
So, the simple \emph{groums} give implicitly more weight to each node and the other edge which are more important for API usage comprehension.
\\
\\
Note that the simplification of the \emph{groums} was done after the detector generation. During the generation process, we consider the data dependencies.

\textbf{\underline{Study 1.C: \textit{with vs. without clustering}.}}

The clustering allows achieving much better accuracy than without clustering.
The clustering benefits the detection accuracy as we can observe on the top 10\% an increase of 3.33\% with the parallel evolution and an increase of 20\% with global evolution on \http whereas \crypto sees an increase of 10\% with the global evolution to obtain 60\% of accuracy (\tabref{tab:process0}, \tabref{tab:process4} and \tabref{tab:process5}).

As we conjectured, the clustering allows to target specifically different usage scenarios during the detector generation. When the detectors are generated without clustering, some usage scenarios can be partially ignored in the random generation of detectors, especially those that are not very common, i.e., the probability to generate a detector from this rare usage scenarios is low.
\\
\\
In conclusion With this first study, we first show that global evolution is more beneficial than parallel evolution because of the presence of small clusters. Second, simple \emph{groums} are slightly more efficient because they give more weight to important information about the API usages. Finally, not performing the clustering is detrimental to the detection accuracy and confirms the intuition that grouping the API usages to generate the detectors is beneficial.

\subsection{Efficiency Analysis}
\label{sec:eval-efficiency}

In this experiment, we aim to answer \textbf{RQ2.} \RQtwo

\subsubsection{Analysis method}

We measure the execution times for different steps of the approach as
well as the average required storage for the 3 APIs when serializing the detectors' \emph{groums}.
We run with the best detector generation strategy which uses simple \emph{groums}
(without data dependencies), clustering and global evolution of detectors for each API
to compute the average performances.
For each execution we compute the time at each step of the detectors generation
and risk-score computation.
For memory usage, we use JVM Monitor~\footnote{http://jvmmonitor.org/index.html} to obtain realtime data.
And for storage, we look at the Windows explorer properties of the output folders.
To measure the performance of our approach, we use the
benchmark \emph{MuBench}~\cite{MuBench}.

\tool{} works in 2 steps, the detector generation and the risk evaluation.
The first step takes relatively a long time with in average 33 minutes to generate the detectors for the 3 APIs.
The generation of detectors consists of 3 steps: extracting API \emph{groums} from source code, clustering
usages and generating the detectors. The extraction of APIs \emph{groums} takes less than 10 minutes and varies
among APIs depending on the number of methods provided as input. Clustering takes less than 15 seconds.
As for the detectors, the generation exceeds 15 minutes to produce 50 detectors per API.
Note that the detector generation process is performed once and the generated detectors can be used several time to evaluate different clients using the API for which the detectors were generated. The minutes magnitude can be acceptable.

We measured the execution time of the risk estimation only for \Iterator and \crypto because MuBench has not client code and misuses for \http. The risk estimation is also divided into 3 steps which are listed in Table 5. The first step is the extraction of API \emph{groums} from client code under evaluation. This step takes less than 23s for \Iterator and less than 10s for \crypto. This difference in time is explained by the fact that \Iterator clients compose a corpus of more than 21,000 methods and produce 2,175 uses. In contrast, \crypto has only 20 different uses in 120 methods. The second step is to load the trained detectors to use them for the detection. This takes about 10s for 50 detectors for each of the two APIs. Finally, there is the sorting time which are less than 9 seconds for \Iterator and about a tenth of a second for \crypto. This is explained by the difference in the number of methods to sort, 100 times more for \Iterator.

For memory measurement, we ran the experiment
for the 3 APIs with clustering and global evolution on a laptop under Windows
10 with an OS architecture amd64, 8 processors and 6 GB of RAM.
On average, an API \GROUM{} is stored on 26 MB of memory and a detector on 2.5
MB. It is, therefore, necessary to provide more than 125 MB of memory to save
50 detectors.
In terms of heap memory the maximum is 3,695 MB and on average it is used 1,109 MB.

In conclusion, \tool{} requires more time to generate the detectors, but this is done only once. It requires much less time to evaluate client programs and assign risk scores to their methods. This short time allows potential integration into IDEs.

\begin{table}[htp]
	\centering
	\caption{Evaluation time in seconds on MUBench clients.}
	\begin{tabular}{|l|r|r|}
		\hline
		 						 	& Iterator  & crypto \\
		\hline
		API groums building &  24.64 &  9.41 \\
		%\hline
		Detectors deserialization &  10.34  &  10.15 \\
		%\hline
		Ranking time &  8.81  &  0.16 \\
		\hline
		Total time &  43.80 &  10.31 \\
		\hline
	\end{tabular}
  \label{tab:ExecTime}
\end{table}

\subsection{Comparative Analysis}
\label{sec:comparison}

In this experiment, we aim to answer \textbf{RQ3.} \RQthree

\subsubsection{Analysis method}

To answer this research question, we run \tool to detect misuses in MuBench
dataset~\cite{MuBench} and compare the result with those of the misuse
detection tools in MuBenchPipe~\cite{MuBenchPipe}. MuBench provides a
benchmark of API misuses from real-world projects, which have been manually
verified. MuBenchPipe provides a pipeline for running existing detectors on
the misuses in MuBench. In the current version, MuBenchPipe supports four
Java API misuse detectors: DMMC~\cite{MBM10},  GROUMiner~\cite{Grouminer},
Jadet~\cite{WZL07}, and Tikanga~\cite{WZ11}, all of which mine patterns from
each subject project and detect misuses as violations of the patterns at the
same time.
MuBench contains misuses from a wide range of APIs including \code{Iterator}
and \code{javax.crypto}. However, it does not contain misuses from
\code{javax.serlet.http}. Therefore, in this experiment, we compare the detectors on only misuses
of \code{Iterator} and \code{javax.crypto}. MuBench has 13 and 7 misuses for respectively
\code{Iterator} and \code{javax.crypto} , .

Using the detectors trained from our dataset collected from GitHub, we
detected misuses in the methods of the projects in the benchmark and ranked
the analyzed methods according to their risk scores. For \code{Iterator}
uses, we only considered the top-13 ranked methods, since MUBench just
flagged 13 misuses of Iterator. And for the same reason, we considered the
top-7 ranked methods for \code{javax.crypto}. To compare the tools we calculate
the recall as the number of misuses detected by a tool over the total number
of known misuses in MUBench.
We also went through the identified misuses to qualitatively and conceptually
compare our approach and the considered tools to show how they could
complement each other.

\subsubsection{Results for RQ3}

As shown in Table~\ref{tab:recall}, \tool{} performs better than the others four tools for \crypto and less good than two out of the four tools for \Iterator. What is interesting in the results of our approach is the diversity of misuse revealed. For \crypto we detect 3 misuses of different types, a missing call, a missing exception handling and a missing condition value or state. For \Iterator, all misuses are missing condition value or state. Three considerations may explain the good performance for \crypto.

The first important point is that we do not mine and use usage patterns.
Using patterns restrict the detection to specific scenarios. That focus on
specific usages and do not take into account the diversity of API utilization,
especially rare usages.

Second, our groums capture various control structures, including exceptions, which allows to better represent
the API good usage and the potential deviances. This is why we were able to detect a missing exception.

The last benefit of our approach is the edge typing of \emph{groums}. This allows, among others,
to define the scopes in which the API methods are called. For example, with the loop inclusion edge, it is possible
to distinguish between an API method call inside a loop and the same call after a loop.
This bring more precise information about the usage scenario.
In conclusion, \tool{} can complement the pattern-based detection tools. These are good in detecting specific cases of misuses
as it was the case for \Iterator. However, for diverse misuses, as for \crypto,
\tool{} is better suited as it learns various deviances from normal uses rather than encoding a fixed set of patterns.

\begin{table}[htp]
	\centering
	\caption{Detectors recall (number of flagged misuses)}
	\begin{tabular}{|>{\raggedright}m{2.3cm}|*{6}{m{0.7in}|}}
		\hline
		
		 Tool			& Iterator  & crypto \\
		\hline
		
		\tool	 		&  8\% (1)	&  	 43\% (3)	\\
		\hline
		DMMC 			&  14\% (2) 	&   0\% (0) \\
		\hline
		GROUMiner 		&  0\% (0) 	&   0\% (0) \\
		\hline
		Jadet 			&  0\% (0)	&   0\% (0) \\
		\hline
		Tikanga			&  23\% (3) 	&   0\% (0) \\
		\hline
	\end{tabular}
	
  \label{tab:recall}
\end{table}
 
\section{Discussion and Threats to Validity}
\label{sec:validity}

It is widely recognized that several factors can bias the validity of empirical studies. We will discuss the different threats that may affect our study.
External Validity concerns the possible biases related to the choice of experimental subjects/objects.  Although we have analyzed a large-scale dataset of 4,869 API groums with 8,641 API uses, collected from 6,364 commits of 1,874 projects selected in an initial set of 21,745 repositories. We cannot claim that our results can be generalized beyond the three APIs for which the dataset was collected.
Regarding the Internal validity, we did not fine-tune the other detectors our self,  we rather used the results reported in MUBench with the configurations reported in their respective publications of the different detectors. Probably a better fine-tuning would have lead to different results. Another threat to internal validity can be related to the knowledge and expertise of the human evaluators who compared the groums of the methods before and after the bug fixing commit; To add the method before the fix to the set of misuses and the method after the fix to the set of good uses of the corresponding API. However, it is only the expertise of the human evaluators that guaranty that the result of the fix itself is a good use of the API.
Threats to construct validity can be related to the measurements performed to address the research questions. In our study, we measured the risk score for each evaluated method. The misuse detectors are applied to each good and bad use case contained in the test set. The cases are then sorted by their risk scores. This allows computing the accuracy on the top-ranked cases. We computed the accuracy as the number of true positives (misuses) over the number of considered top-k use cases. While the adoption of these measurements is popular to assess the efficiency of algorithms and to conduct comparative studies we cannot neglect the existence of a slight bias related to the set of ground truth used to calculate accuracy, as the good and bad use case were manually validated. Moreover, the experimenter expectancy effect is another possible threat. Indeed, the manual inspection was performed by two of the authors.  

Our approach does not outperform others in all cases. We have rather developed an approach to complement existing work. Existing misuse detection techniques are mainly based on usage patterns. This strategy is effective in detecting specific cases of misuses in particular on APIs similar to the Iterator where the uses are easy to characterize and patterns are easy to infer. Our approach brings a new way of addressing the API misuse problem. while the vast majority of existing detectors consider the deviance from perfection as a criterion to identify API misuses, we rather decided to opt for the closeness to evil as a criterion to detect the API misuses.  Instead of measuring how far from right behavior (i.e. the patterns) is the evaluated method, we look at how similar is an evaluated method to bad behaviors (i.e. the detectors). Moreover, in our metaphor with the immune system, a cell that tends to match several T-cells would be qualified as pathogenic. Thus, we made the choice to assign higher risk scores to methods that are close to multiple detectors through the logical function of Noisy or, rather than looking at the detector that best matches the method being evaluated. To improve APIMMUNE in detecting specific cases of misuses, as it was the case for Iterator, a different strategy to measure the risk score could be investigated, for instance the maximum similarity between the evaluated method and each of the detectors.  However, for diverse misuses, as for crypto, APIMMUNE can complement the pattern-based detection tools. as it learns various deviances from normal uses rather than encoding a fixed set of patterns.

In our approach, we generate only fifty detectors for each API, this could be considered a low number and we could consider producing a larger number of detectors. However, if we generate a high number of detectors this will bring into the picture the detectors diversity issue. When the detectors are not as diverse as expected, the misuse detection step will be impacted,  If a method (a cell) tends to match several detectors  (T-cells) it would be qualified as misuse (pathogenic). Thus,  we assign higher risk scores to methods that are close to multiple detectors through the logical function of Noisy or. As a consequence, if the detectors are not different from each other we will assign higher risk scores to methods that are similar to the same redundant detectors. Moreover, our choice to limit the number of detectors is also motivated by performance concern. The genetic-based generation of detectors is a heavy process during which thousands of detectors are generated and are compared to the training data set as well as to each other, to finally come up with a limited set of best and most diverse detectors. Despite all those reasons, we have to admit that fixing the number of detectors to allow number for all the evaluated APIs, is a strategy that needs to be improved. in our future work, we will investigate the impact of correlating the number of detectors to some characteristics of the API such as the number of public methods or the number of inferred clusters (usage scenarios of the APIs)."

 Our approach generates a fixed number of detectors that represent artificial signatures that are different from those of the safe code. However, the fact that a detector deviates from the normal API usages in the learning dataset, doesn't guarantee that the detector actually represents by itself a misuse of the API. Moreover, in a very extreme case, we may consider the risk that the detector generation could end up producing signatures that are good or efficient usage of the API. Even though this case is almost impossible, our approach has the features that allow avoiding it. What actually makes a detector deviate from the good usage of an API is the mutation operator. Each detector undergoes one or multiple mutations among nine types of mutations. Moreover, we replace the boolean detection results (Self/non-Self) with a risk score that allows ranking the evaluated methods according to the estimated risk. We assign higher risk scores to methods that are close to multiple detectors through the logical function of Noisy or, rather than looking at the detector that best matches the method being evaluated. Thus even though a bad detector ends up in the list of used detectors, its impact will be reduced through the comparison with other detectors. In addition, the obtained detectors will have their own independent life cycles. They can be reused/shared, enhanced with new detectors when new safe clients are considered and destroyed if they detect false positive(s).

\section{Related Work}
\label{sec:related}

In this section, we focus on presenting the related work on API misuse
detection approaches that use the consensus principle. Those
approaches typically follow two steps: mining the (good) usage
patterns from a given code corpus and considering uses deviating from
the patterns as misuses. Amann \etal \cite{MuBenchPipe}
have recently conceptually compared the capabilities of existing API-misuse
detection approaches, and conducted several empirical studies to evaluated
and compared four well-known detectors using a benchmark
infrastructure called MUBench~\cite{MuBench}. The authors reported the
existing API misuse detectors have suffered high false positives due
to an important issue with the use of thresholds on frequent usages
(patterns) and on the deviations from those patterns.

JADET is a misuse detector for Java~\cite{WZL07} that focuses
on API call order and call receivers in usages. It derives a pair of
calls for each call order. The sets of these pairs are the inputs to
the mining to identify API patterns, in term of {\em sets of pairs of
  API calls}. JADET is capable of detecting missing method calls and
missing loops as a missing call-order relation from a method call in
the loop header to itself. Tikanga~\cite{WZ11} is built up on JADET,
and extends it to general Computation Tree Logic formulae on
object usages with the use of model checking on those. Tikanga
applies Formal Concept Analysis~\cite{GW99} to mine patterns and
misuses at the same time.

Tapir \cite{saied2020towards, saied2018towards} considers the recovery of temporal API usage patterns as an optimization problem and solves it using a genetic-programming algorithm.  API temporal constraints are mined from execution traces of client programs using the API, and  Linear Temporal Logic (LTL) formulas, representing candidate usage patterns are recovered. Tapir alert the user of potential API misuse of when LTL formulas, are violated.

Nguyen {\em et al.}~\cite{Grouminer} propose a graph-based object
usage model (groum) for their misuse detector in GrouMiner. It uses
sub-graphs mining to mine patterns as frequent usage
sub-graphs. GrouMiner can detect misuses with missing API elements
compared to patterns. In our work, we do not abstract/learn usage
patterns from historical data to use them for the detection. The
novelty of our approach is in generating new artificial data, i.e.,
detectors, that diverge from the historical data, following the
negative selection principle. By doing this, we create new artifacts
(detectors) that can be shared among development groups without
disclosing the clients' code.

DMMC~\cite{MBM10,MM13} aims to detect misuses in API method calls with
exactly one missing one. It focuses on type usages, i.e., sets of
methods called on a given receiver type of a given method. The
assumption is that violations should have only few exactly similar
usages, but many near-similar ones.
Alattin~\cite{TX09b} mine alternative patterns for condition checking.
It applies frequent-itemset mining on the set of rules on pre- and
post-condition checks on the receiver, the arguments, and the return
value of a method call. It detects missing \code{null}-checks and
missing value or state conditions that are ensured by checks.
CAR-Miner~\cite{TX09} aims to detect API misuses in error handling.
To detect a misuse, it extracts the normal call sequence and the
exception call sequence for a method call. It learns the association
rules and then determines the expected exception handling and reports
a violation if the actual sequence does not include. CAR-Miner can
detect missing exception-handling as well as missing method calls
among error-handling functions.
AX09~\cite{AX09} also detects incorrect error handling. It uses model
checking to generate error handling paths as sequences of method calls
and applies frequent-subsequence mining to detect patterns. It then
uses push-down model checking to verify the consistency to these
patterns and identify respective misuses. 

PR-Miner~\cite{LZ05} encodes API usages as a set of all function names
in a function in C, and uses frequent-itemset mining to detect
patterns. Misuses are the subset that occurred at least 10 times less
frequently than the pattern. It focuses on detecting missing method
calls.
Chronicler~\cite{RGJ07b} aims to detect patterns in frequent call
orders in inter-procedural control-flow graphs. The orders hold at
least 80\% of the all execution paths are patterns and otherwise, they
are violations. Chronicler detects missing method calls.  Since loops
are unrolled exactly once, it cannot detect missing iterations.
DroidAssist~\cite{NPVN16} detects misuses in Java bytecode. It learns
the call sequences to build a Hidden Markov Model. If a likelihood
of a given call sequence is too small, it is considered as a misuse.
MUDETECT \cite{rendemystify} increases the level of details found in identified patterns, in order to increase the accuracy and recall of API misuse detection. MUDETECT  uses a  new graph representation of  API  usages that captures different types of  API  misuses and it devised a systematically designed ranking strategy that effectively improves precision. More recently Ren et al. \cite{sven2019investigating} devised a text mining technique that extracts Android API misuses and patches from StackOverflow answers. The method produces a natural language report from these code fragments, explaining how to use the API.

\begin{table*}[t]
	\centering
	\caption{Comparison of different misuse detection tools}
	\begin{tabular}{|l|l|l|l|}
		\hline
		
		 Tools			& Mono API detection  & Input  to learning step &  Type of detected misuse \\
		\hline
		
		Alattin~\cite{TX09b}	 	&No	& Code examples extracted & Missing null checks \\
		                            &   & using code search engines	&  Missing condition value or state \\
		\hline
		AX09~\cite{AX09}  	&No		&  Client systems source code 	&   Incorrect error handling \\
		\hline
		CAR-Miner~\cite{TX09} 	&No	&  Code examples extracted 	&   Incorrect error handling \\
		  &   & using code search engines	&   \\
		\hline
		Chronicler~\cite{RGJ07b} 	&No		&  Client systems source code	&   Missing method call \\
		\hline
		DMMC~\cite{MBM10,MM13} 	&No		&  Java byte code of Client systems  	&   Missing method call \\
		\hline
		DroidAssist~\cite{NPVN16}	&Yes		&  Java byte code of Client systems 	&  Missing method call \\
		\hline
		Jadet~\cite{WZL07} 	&No		&  Client systems source code 	&   Missing method calls  \\
		&&& Missing loops\\ 
		\hline
		PR-Miner~\cite{LZ05}	&Yes		&  Client systems source code	&  Missing method calls  \\
		\hline
		Tikanga~\cite{WZ11}	&No		&  Client systems source code 	&   Missing condition value or state \\
		\hline
		GrouMiner~\cite{Grouminer} 	&No		&  Client systems source code 	&   Missing API elements \\
		\hline
	\end{tabular}
	
  \label{tab:related}
\end{table*}

As we can see in table \ref{tab:related}, in the majority of the cases the misuse detection is not specific to a single API and thus it fails in identifying different types of misuses related to that API.  Moreover, the learning step is dependent on the considered client systems that generally focus on specific usages and do not take into account the diversity of API utilization, especially rare usages

\section{Conclusion}
\label{sec:conclusion}

In this work, we propose \tool{}, an approach to detect API misuses using the immune-system metaphor.  The normal API usages are considered as self normal body cells. Whereas API misuses are considered as the non-self antigens. We use a genetic algorithm to generate detectors mimicking the T-cells that can be used to detect non-self  API misuses. This approach has the advantage of generating the detectors once, and then, they can be used and enhanced without the need of disclosing the clients’ code, nor abstracting good use patterns. Moreover, the detectors can be produced for different versions of the programming interface, which brings more flexibility to the detection process.  The evaluation of \tool{} shows that it can detect various types of misuses, however, our approach does not outperform others in all cases.  We have rather developed an approach to complement existing work. Existing misuse detection techniques are mainly based on usage patterns. This strategy is effective in detecting specific cases of misuses in particular on APIs where the uses are easy to characterize and patterns are easy to infer. Moreover, pattern-based approaches are heavily dependent on the frequency of good usage in the learning data set. \tool{} can complement the pattern-based detection tools that are limited to fixed sets of misuse patterns. For diverse misuses, \tool{} is better suited as it learns various deviances from normal uses rather than encoding a fixed set of patterns. \tool{} may require some time to generate the detectors, but this is done only once. It requires much less time to evaluate client programs and assign risk scores to their methods. This short time encourages potential integration into IDEs. Despite the encouraging results, there is still some exploration to be done with \tool{}  the approach has multiple parameters that had to be set. It could be interesting to explore the results with different threshold parameters to see their impact on detection efficiency. For instance, we could increase the number of generated detectors while taking into account the potential redundancy of generated detectors. In the future, we plan to experiment with larger datasets involving many APIs. This will help us customize the detection process to the characteristics of these APIs (number of public methods, the number of distinct functionalities, etc.). Another area for improvement is to explore the combination of our approach with pattern-based detection.

\bibliographystyle{IEEEtran}
\bibliography{bibliography,api-refs}

\vspace{12pt}

\end{document}